\documentstyle[12pt]{article}
\def\lb       {\left( }
\def\rb       {\right) }

\def\lbb     {\left[ }
\def\rbb      {\right] }
\def\comma      { \, , }
\def\period     { \, . }
\def\bra#1      { \langle \, #1 \, \vert \, }
\def\ket#1      { \, \vert \, #1 \, \rangle \, }
\def\semiket#1  { \, #1 \, \rangle \, }
\def\del        {  \partial  }
\def\half       {  {1\over 2}  }

\def\abs#1      {  \, \vert #1 \vert \,   }
\def\Im#1    { \, {\rm Im } \, #1  }
\def\Re#1    { \, {\rm Re}  \, #1  }

\def\bfR     { {\bf R}}
\def\bfZ     { {\bf Z}}

\def\vecii#1#2      {  \left(\begin{array}{c}#1\\#2\end{array}\right)  }
\def\veciii#1#2#3   {  \left(\begin{array}{c}#1\\#2\\#3\end{array}\right)  }
\def\matrixii#1#2#3#4            {  \biggl( \begin{array}{cc}#1&#2\\#3&#4
                                       \end{array} \biggr) }
\def\matrixiii#1#2#3#4#5#6#7#8#9 {  \left(\begin{array}{ccc}#1&#2&#3\\
                                     #4&#5&#6\\#7&#8&#9\end{array}\right)  }
\def\eqabegin         {  \begin{eqnarray}  }
\def\eqaend           {  \end{eqnarray}  }
\def\nn               {  \nonumber  }

\def\sectionnumbering { \setcounter{equation}{0}
         \renewcommand{\theequation}{\arabic{section}.\arabic{equation}}}

\def\mysection#1{ \addtocounter{section}{1} \setcounter{subsection}{0}
                 \sectionnumbering 
      \begin{center} {\sc \arabic{section} \quad  #1 } \end{center}   
    \par \bigskip}

\def\csectionast#1    { \begin{center}  
    {\large\bf #1  }   \end{center} \par \bigskip}
\renewcommand{\thefootnote}{\fnsymbol{footnote}}

\def\xxx#1 {{\tt hep-th/#1}}
\def\grqc#1 {{\tt gr-qc/#1}}
\def\npb#1(#2)#3 { Nucl. Phys. {\bf B#1} (#2) #3 }
\def\rep#1(#2)#3 { Phys. Rept.{\bf #1} (#2) #3 }
\def\plb#1(#2)#3{Phys. Lett. {\bf #1B} (#2) #3}
\def\prl#1(#2)#3{Phys. Rev. Lett.{\bf #1} (#2) #3}
\def\prd#1(#2)#3{Phys. Rev. {\bf D#1} (#2) #3}
\def\ap#1(#2)#3{Ann. Phys. {\bf #1} (#2) #3}
\def\rmp#1(#2)#3{Rev. Mod. Phys. {\bf #1} (#2) #3}
\def\cmp#1(#2)#3{Comm. Math. Phys. {\bf #1} (#2) #3}
\def\mpl#1(#2)#3{Mod. Phys. Lett. {\bf A#1} (#2) #3}
\def\ijmp#1(#2)#3{Int. J. Mod. Phys. {\bf A#1} (#2) #3}
\def\mpla#1(#2)#3{Mod. Phys. Lett. {\bf A#1} (#2) #3}
\def\jhep#1(#2)#3{JHEP {\bf  #1} (#2) #3}
\def\cqg#1(#2)#3{Class. Quant. Grav {\bf  #1} (#2) #3}
\def\delbar {\bar{\partial}}
\def\zbar   {\bar{z}}
\def\kket#1   { \vert #1 \rangle \! \rangle }
\def\bbra#1   { \langle \! \! \langle  #1 \vert }

\def\Dlw      { {\cal D}_{\rm lw} }
\def\Dhw      { {\cal D}_{\rm hw} }
\def\Dpc      { {\cal D}_{\rm pc} }
\def\gammabar  { \bar{\gamma} }
\def\xbar     { \bar{x} }
\def\betabar     { \bar{\beta} }
\def\mbar     { \bar{m} }
\begin{document}
%
\def\papertitlepage{\baselineskip 3.5ex \thispagestyle{empty}}
\def\preprinumber#1#2#3{\hfill \begin{minipage}{4.2cm}  #1
              \par\noindent #2
              \par\noindent #3
             \end{minipage}}
\renewcommand{\thefootnote}{\fnsymbol{footnote}}
%
%
\papertitlepage
\setcounter{page}{0}
\preprinumber{December 1999}{UTHEP-417}{hep-th/0005169}
\baselineskip 0.8cm 
\vspace*{2.5cm}
\begin{center}
{\large\bf On string theory in $AdS_3$ backgrounds\footnote{
  Talk given at YITP workshop `Developments in Superstring and M-theory',
  Kyoto, Japan, October 27-29, 1999. }}
\end{center}
\vskip 10ex
\baselineskip 1.0cm
\begin{center}
        {\sc Yuji  ~Satoh}  \\
 \vskip 3ex
    {\sl Institute of Physics, University of Tsukuba} \\
 \vskip -2ex
   {\sl Tsukuba, Ibaraki 305-8571, Japan} \\
 \vskip -2ex
   {\tt ysatoh@het.ph.tsukuba.ac.jp}
\end{center}
\vskip 14ex
%
\baselineskip=3.5ex
\begin{center} {\bf Abstract} \end{center} 

We discuss the string theory on $AdS_3$.
In the first half of this talk, we review the $SL(2,R)$  
and the $SL(2,C)/SU(2)$ WZW models which describe the strings 
on the Lorentzian and Euclidean $AdS_3$ without RR backgrounds, respectively.
An emphasis is put on the fundamental issues such as the unitarity, 
the modular invariance and the closure of the OPE. 
In the second half, we discuss some attempts at clarifying such problems.
In particular, we discuss the modular invariance of the $SL(2,R)$ WZW 
model and the calculation of the correlation functions of the 
$SL(2,C)/SU(2)$ WZW model using the path-integral approach.
%
%
%
%
%
%
%
\newpage
\renewcommand{\thefootnote}{\arabic{footnote}}
\setcounter{footnote}{0}
\setcounter{section}{0}
\baselineskip = 0.6cm
\pagestyle{plain}
%
\mysection{Introduction : why string theory on $AdS_3$}
%
%
In this talk, I would like to discuss the string theory on 
$AdS_3 $, namely, $SL(2,R)$ or its Euclidean analog $ SL(2,C)/SU(2)$. 
Besides the recent intensive studies, 
this string theory has in fact been investigated for 
more than a decade from various interests \cite{BFOW}-\cite{Teschner1}. 

First of all, string theory in backgrounds with curved time is not well 
understood. There are 
several such models which are relatively well studied \cite{crvdstrng}, 
but the analysis is essentially reduced to that of the free theory.
In this respect, the string theory on $SL(2,R)$ seems to give the simplest 
truly interacting model. 
This is because $SL(2,R)$ is a very simple space-time with
maximal symmetries and the corresponding model is described by the 
$SL(2,R)$ WZW model when there are no RR charges.

Second, related to the above, little is known about non-compact 
(non-rational) CFTs \cite{Gawedzki}. 
Again, the $SL(2,R)$ WZW model or its Euclidean analog, $SL(2,C)/SU(2)$ 
WZW model, gives the simplest one.

Third, it is known that the $SL(2,R)$ WZW model is closely related to 
the string models in various black hole backgrounds. For instance,
the $SL(2,R)/U(1)$ WZW model, which is obtained by a coset, describes
the strings in two-dimensional black hole backgrounds \cite{sl2/u1}. 
An orbifold of the $SL(2,R)$ WZW model \cite{3dbh,NS} 
gives the string model 
in the three-dimensional BTZ black hole geometry \cite{BTZ}. 
When a five-dimensional black hole corresponding to the D1-D5 system
is lifted to six dimensions, its near-horizon geometry becomes 
$AdS_3 \times S^3$ \cite{HSS} 
(precisely speaking (BTZ black hole)$ \times S^3$). By further taking 
an S-dual, the system is described by the $SL(2,R) \times SU(2)$ WZW
\cite{CT}. Similarly, $AdS_3$ or the BTZ black hole
appears quite generally as the near-horizon geometry of the black strings 
obtained by lifting charged black holes in generic dimension \cite{YS2}.

Finally, closely related to the above D1-D5 system, 
the string theory on $AdS_3$ gives the simplest 
case of the AdS/CFT correspondence \cite{AdSCFT}. This aroused 
the renewed interest and many works have been devoted to the study 
of the strings on $AdS_3$
in the cases both with \cite {GS,BVW} and without \cite{GKS}-\cite{BDM} 
RR charges.

However, in spite of recent progress, it seems that there still remain
open questions about the string theory on $AdS_3$ itself 
at the fundamental level. Such a state of the problem was recently discussed 
in \cite{PMP}.
In this talk, we will focus on the cases without RR charges.
In the next two sections, 
we will review the $SL(2,R)$ WZW model
and its Euclidean analog, the $SL(2,C)/SU(2) = H_3^+$ WZW model.
We will see that our understanding 
is still incomplete on the fundamental consistency conditions of string theory 
such as the unitarity, the modular invariance and the closure 
of the operator product expansions. Hence we will discuss some attempts 
towards better understanding in the following sections. 
In section 4, we will discuss modular invariance of the $SL(2,R)$ WZW
model and obtain some important information about the spectrum \cite{KYS}. 
In section 5, we will discuss
the calculation of the correlation functions of the $H_3^+$ WZW model
using a path-integral approach \cite{IOS}.
We will conclude with a brief summary.
\par \vskip 1ex \noindent
\mysection{$SL(2,R)$ WZW model}
%
%

Let us start with the discussion of Lorentzian $AdS_3$. It 
is defined by the following metric and the embedding equation,
\eqabegin
   ds^2 &=& -dx_0^2 - dx_1^2 + dx_2^2 + dx_3^2 \comma \nn \\
   -l^2 &=& - x_0^2 - x_1^2 +  x_2^2 + x_3^2 \period \label{AdS3}
\eqaend
This is a maximally symmetric space with negative constant curvature 
and a solution to the three-dimensional Einstein's equations with
a negative cosmological term $-l^{-2}$,
\eqabegin
   R_{\mu \nu} &=& -2 l^{-2} g_{\mu \nu}
  \period
\eqaend

The space-time defined in the above is 
the same as the group manifold $SL(2,R)$. Hence without RR charges
the (bosonic) string theory in this background is described by 
the $SL(2,R)$ WZW model. Its action is given by 
\eqabegin
   S &=& - \frac{k}{8\pi} \int_{\Sigma} {\rm Tr} \bigl( d g dg^{-1} \bigr)
      + \frac{ik}{12\pi} \int_{B} {\rm Tr} \bigl( g^{-1} dg \bigr)^3
   \comma \label{WZW}
\eqaend 
where $g(z) \in SL(2,R) $, $k$ is the level, $\Sigma$ is a two-dimensional
surface (world-sheet) and $B$ is a three-dimensional manifold satisfying
$\del B = \Sigma$. The action has the 
$\hat{sl}(2,R)_L \times \hat{sl}(2,R)_R$ current algebra symmetry. The 
corresponding currents are 
\eqabegin
   J(z) = \frac{ik}{2} \del g g^{-1} \comma && 
   \tilde{J}(\bar{z}) = \frac{ik}{2} g^{-1} \delbar g 
  \period
\eqaend 
Here we have denoted the quantity in the right sector by tilde.
In the following we will omit the expressions in the right sector
unless they are necessary.
The model has the conformal symmetry and 
its  energy-momentum tensor is given by the Sugawara form,
\eqabegin
   T(z) &=& \frac{1}{k-2} \eta_{ab} J^a(z) J^b(z)
   \comma
\eqaend
where  $\eta_{ab} =$ diag $(-1,1,1)$. $J^a(z)$ are defined through
$J(z) = \eta_{ab} \tau^a J^b (z)$ with $\tau^a \in sl(2,R)$.
In terms of the modes of $J^a(z)$, those of $T(z)$ are written as
\eqabegin
  L_n & = & \frac{1}{k-2} \sum_{m \in \bfZ } 
   : \half J^+_{n-m} J^-_{m} + \half J^-_{n-m} J^+_{m}
     - J^0_{n-m} J^0_{m}: 
   \period 
\eqaend
These currents and the energy momentum tensor satisfy the 
following commutation relations
\eqabegin
 \lbb J^0_n , J^0_m \rbb & = & - \half  k n \delta_{n+m} \comma  \qquad
 \lbb J^0_n , J^\pm_m \rbb \ = \  \pm J^\pm_{n+m} \comma  \nn \\
 \lbb J^+_n , J^-_m \rbb & = & - 2 J^0_{n+m} + k n \delta_{n+m}
     \comma \nn \\
  \lbb L_n , J_m^a \rbb &=& -m J^a_{n+m} \comma  \\
  \lbb L_n , L_m \rbb &=& (n-m) L_{n+m} + \frac{c}{12} n(n^2-1) \delta_{n+m}
  \comma \nn
\eqaend
with $c$ being the central charge given by
\eqabegin
  c &=& \frac{3k}{k-2} \period
\eqaend

Because of the symmetry, the states at the lowest grade are classified 
by the representations of $SL(2,R)$. They are  labeled by the values 
of $J^0_0$ and  
$ \vec{J}^2 = \frac{1}{2} J^+_{0} J^-_{0} + \frac{1}{2} J^-_{0} J^+_{0}
     - J^0_{0} J^0_{0} $. These operators act as
\eqabegin
   \vec{J}^2 \ket{ j,m } = -j(j+1) \ket{ j,m } \comma &&
   J^0_0 \ket{ j,m } = m \ket{ j,m } \period
\eqaend
Since $-j(j+1)$ is invariant under $ j  \to -j-1 $, one can always
bring the values of $ j $ 
into the region Re $ j \leq -1/2 $ and Im $ j \geq 0$.
We will take this convention.  A generic state in the left sector 
is obtained by acting on $ \ket{ j, m } $ with $J^a_{-n}$ $(n \geq 0)$
and takes the form 
\eqabegin
   \bigl( J_{-n_1}^{a_1} J_{-n_2}^{a_2} \cdots \bigr) \ket{ j, m }
   \period \label{leftst}
\eqaend
A generic states in the model is obtained by tensoring (\ref{leftst})
and a similar expression of the right sector.  

Since we expect the model to be  unitary, we choose the unitary 
$SL(2,R)$ representations for the zero-mode part. 
There are five classes of such representations.
For the universal covering group of $SL(2,R)$, they are 
\par\medskip\noindent
\hspace*{0.4em} 
(1) \  Identity representation $ {\cal D}_{\rm id} $ : 
   the trivial representation with $\vec{J}^2 = J^0_0 = 0$.
\par\noindent 
\hspace*{0.4em} 
 (2) \ Principal continuous series $\Dpc$: representations 
     with $ m = m_0 + n, \, 0 \leq m_0 < 1$,   $ n \in \bfZ $ 
   \\ \hspace*{2.4em}  
     and $ j = -1/2 + i \rho, \, \rho > 0 $.
\par\noindent
\hspace*{0.4em} 
 (3) \  Supplementary series ${\cal D}_{\rm sup}$: representations  
     with $ m = m_0 + n, \, 0 \leq m_0 < 1, \, n \in \bfZ $  and 
  \\ \hspace*{2.4em}
   min$\{ -m_0, m_0-1 \} < j \leq -1/2 $.
\par\noindent
\hspace*{0.4em} 
 (4) \  Highest weight discrete series $\Dhw$ : representations 
      with $ m = M_{\rm max} - n$, $ n = 0,1,2,...,$ 
 \\ \hspace*{2.4em}
   $ j = M_{\rm max} \leq -1/2$ and 
 the highest weight state satisfying 
   $J^+_0 \ket{ j,j } = 0$.
\par\noindent
\hspace*{0.4em} 
 (5) \ Lowest weight discrete series $\Dlw$: representations 
    with $ m = M_{\rm min} + n $, $ n = 0,1,2,...,$ 
   \\ \hspace*{2.4em}
        $ j = - M_{\rm min} \leq -1/2$ 
  and the lowest weight state satisfying 
  $J^-_0 \ket{ j,-j } = 0$.

\medskip \noindent
If we do not take the universal covering group, the parameters are restricted 
to $m_0 = 0, 1/2$ in (2),
$ m_0 = 0 $ in (3) and $ j = $ (half integers) in (4) and (5). 

The harmonic analysis on $SL(2,R)$ shows that the 
square-integrable functions are decomposed into the 
representations of $\Dpc$, $\Dhw$ and $\Dlw$. Schematically,
\eqabegin
   L^2 \bigl( SL(2,R) \bigr) & \sim & 
   \sum_{j < -1/2} \bigl(-j-\half \bigr) 
 \bigl( \Dhw ^j \oplus \Dlw^j \bigr) \oplus 
    \int_{0}^\infty  d \rho \, f(\rho) \, \Dpc ^{-1/2+i \rho} 
   \comma
\eqaend 
where $f(\rho)$ is a certain measure (for details, see, e.g., \cite{VK}).
%
\par\bigskip\noindent
{\it Ghost problem}
\par\medskip

Soon after the study of the string theory on $SL(2,R)$ was initiated,
it turned out that the model contains negative-norm physical states, 
namely, ghosts \cite{BFOW}. In the flat case, the original model 
(in the conformal gauge)
also contains negative-norm states because of the time direction. 
However,
the physical state conditions $( L_n - \delta_n ) \ket{ \Psi } = 0 $ 
$(n \geq 0)$ are sufficient to remove such states.  
The result in \cite{BFOW} indicates that this does not work in the $SL(2,R)$
case. In fact, it is easy to find the ghosts.

To see this, we first note the on-shell condition 
\eqabegin
   \bigl( L_0 - 1 \bigr) \ket{ \Psi } = 0 \comma 
   && L_0 = - \frac{j(j+1)}{k-2} + N \comma \label{onshell}
\eqaend
where $N$ is the grade. This means that the spin 
at the zero-mode part, $ \ket{ j, m } $, should be 
\eqabegin
  j &=& j (N) \ \equiv \ - \half \Bigl(\, 1 + \sqrt{1+4(k-2)(N-1)} \, \Bigr)
  \comma 
\eqaend
which corresponds to $\Dhw$ or $\Dlw$ for $k>2$ and $N>1$. Next, we 
consider a set of states 
\eqabegin
   \Bigl\{ J_0^a \cdots J^b_0 \ket{ E_N } \Bigr\} \comma &&
   \ket{ E_N } = (J^+_{-1})^N \ket{ j(N), j(N) } 
   \period
\eqaend
We then find that all the above states are physical but form 
a {\it non-unitary} representation of $SL(2,R)$ for a sufficiently
large $N$. This is because $ \ket{ E_N } $ behaves like a highest weight
state of an $SL(2,R)$ representation with $ j = m = j(N) + N > 0$.
Thus we have found the ghosts. 

Having found that the model contains ghosts, one might think that 
the $SL(2,R)$ WZW model is sick. However, there are several pieces of 
evidence that the model should make sense. First of all, in the weak
curvature limit the model becomes the flat model and hence one should
be able to get a sensible model at least at weak curvature. Second, 
the authors of \cite{BFOW} studied the particle limit of the model
but did not find any pathologies. Third, the effective action of 
the bosonic $\sigma$-model was studied in \cite{FT}. There it was found that 
the effective action has an extremal point corresponding to $AdS_3$
and the model is unitary at one-loop. Finally, as discussed in the 
introduction, the near-horizon geometry of the D1-D5 system is 
described by the $SL(2,R) \times SU(2) $ WZW model after an S-dual
transformation. Since the D1-D5 system is unitary (at least at weak coupling),
we expect that the $SL(2,R) \times SU(2) $ WZW model should also be unitary.

\par\bigskip\noindent
{\it Resolution of the ghost problem}
\par\medskip

The above argument implies that 
we might be missing something important and it might be 
possible to get a sensible theory by finding out an appropriate 
treatment. There are actually two types of the proposals for the
resolution of the ghost problem.

In one proposal \cite{ng1}, the discrete series $\Dhw$ and $\Dlw$ are 
used and the claim is that if we truncate the spectrum so that 
the spin and the level are restricted to 
\eqabegin
  && \half \ \leq \ -j \ < \ \frac{k}{2} \comma \quad k \ > \ 2  \label{ub} 
  \comma
\eqaend
one can remove the ghosts. We call this the unitarity bound.
This bound seems natural if we recall the 
argument of the $SU(2)$ WZW model. In that case, to maintain the unitarity
or the modular invariance, one needs to truncate the $SU(2)$ spin
so that \cite{GO,GW}
\eqabegin
   &&  0 \ \leq \ j \ \leq \ \frac{k}{2} \period 
\eqaend 
Such a truncation is compatible with the closure of the 
OPE and the Ward identities. Thus it is completely sensible. 

However, in the $SL(2,R)$
case, it is not clear if the truncation (\ref{ub}) is compatible with 
other consistency conditions of string theory. 
This is because such consistency conditions are not well 
understood either.(Regarding the discussion of the OPE 
in the Euclidean case, see \cite{Teschner2}.)
Moreover, 
from the on-shell condition (\ref{onshell}), the unitarity bound means
the truncation of the string excitation $N$. This seems 
physically unnatural. In addition,  
the dimensions of the primaries $L_0 = -j(j+1)/(k-2)$
are negative for the discrete series when $k>2$.

In the other proposal \cite{Bars,YS}, the principal continuous series
$\Dpc$ is used.
One way to understand this argument 
is to start with a Wakimoto-like representation
of $\hat{sl} (2,R)$ \cite{Wakimoto} 
using one free boson $\phi$ and the $\beta$-$\gamma$
system. We then bozonize the $\beta$-$\gamma$ by two free bosons.
One of the points there is that an additional zero-mode
is introduced through this bosonization. Here, it may be 
useful to recall that 
the primary states $ \ket{ j, m } $ have only two zero-modes
whereas those in the three-dimensional flat theory  
have three as $ \ket{ p^0,p^1,p^2 } $ (though the 
total zero-modes in the left 
and the right sectors are three in both cases). Thus it seems natural 
to incorporate another zero-mode if we expect that the $SL(2,R)$ model 
smoothly leads to the flat model in the weak curvature limit.
The added zero-mode turn out to give the sector
satisfying 
\eqabegin
    \int \frac{\del \gamma}{\gamma} & \neq & 0
   \comma
\eqaend 
which is called the long string sector \cite{GKS}.
Then by carefully treating the zero-mode part, we find that the on-shell
condition picks up the spins $j=-1/2 + i \rho $ which precisely 
correspond to $\Dpc$. Finally, using the expressions in terms of 
the free bosons, the no-ghost theorem is shown similarly 
to the flat case. In this proposal, the smooth flat limit is 
achieved by taking $k \to \infty$. In addition, the applications to
the black holes discussed in the introduction appear to be 
straightforward \cite{YS}. 

Nevertheless, as in the previous case, it is not clear if this
proposal is compatible with the other consistency conditions.
(For the discussions of the OPE and the modular invariance in this case,
see \cite{BDM} and \cite{YS} respectively.)

In fact, we must say that there is no agreement about 
how to construct the sensible theory of the $SL(2,R)$ strings.
Therefore, to clarify this issue it is very important 
to further investigate the 
fundamental problems such as the modular invariance, the closure of the OPE, 
how to choose the spectrum and how to calculate the correlators.
\par \vskip 3ex\noindent
\mysection{$SL(2,C)/SU(2)$ WZW model}
%
%

In the previous section, we discussed the fundamental open questions
about the $SL(2,R)$ WZW model.
Now let us turn to the discussion of the $SL(2,C)/SU(2) = H_3^+ $
WZW model. The precise formulation of the AdS/CFT correspondence 
requires Euclidean anti-de Sitter spaces \cite{AdSCFT}. 
Euclidean $AdS_3$ is called $H_3^+$ and given by
\eqabegin
   ds^2 &=& -dx_0^2 + dx_1^2 + dx_2^2 + dx_3^2 \comma \nn \\
   -l^2 &=& - x_0^2 + x_1^2 +  x_2^2 +  x_3^2 \period
\eqaend
Note the sign-flips compared with the Lorentzian case (\ref{AdS3}).
This space is also a maximally symmetric space with negative 
constant curvature.

To get a string background, one needs to introduce the NS $B_{\mu \nu}$
field. In some parametrization, the action takes the form,
\eqabegin
   S &=& \frac{k}{\pi} \int d^2 \sigma \Bigl( \del \phi \del \phi \, + \, 
      e^{2\phi} \del \gammabar \delbar\gamma \Bigr)
   \period \label{H3+} 
\eqaend
Here $\gammabar = \gamma^*$ and $\phi \to + \infty $ corresponds to 
the boundary of $H_3^+$. If $\gamma $  and $ \gammabar $ 
are independent, the geometry becomes Lorentzian $AdS_3$.
This action is obtained also by substituting 
$g(z) = h(z)h^\dagger (z)$ with
\eqabegin
  h &=& \matrixii{1}{\gamma}{0}{1} \matrixii{e^{-\phi/2}}{0}{0}{e^{\phi/2}}
    \ \in \ SL(2,C)
\eqaend 
into (\ref{WZW}) \cite{Gawedzki}. 
From this construction, the coset structure $SL(2,C)/SU(2)$ is obvious 
and one finds that the model is actually a WZW model.
%
In terms of $\phi$, $\gamma$ and $\gammabar$,
the functional measure takes a non-trivial form 
\eqabegin
 && {\cal D} \phi {\cal D} ( e^{\phi} \gamma ) {\cal D} 
( e^{\phi} \gammabar) 
  \period \label{measure}
\eqaend

The action has the current algebra symmetry 
$\hat{sl}(2,C) \times \hat{sl}(2,C)^* $. 
In this case, the left and the right 
symmetries are the complex conjugate to each other. The currents of the 
global symmetry acting on the zero-mode part are realized by
\eqabegin
   J^-_0 & = & \del_{\gamma} \comma \qquad 
   J^0_0 \ = \ \gamma \del_\gamma - \half \del_\phi \comma \nn \\
   J^+_0 &=& \gamma^2 \del_\gamma - \gamma \del_\phi - \, e^{-2\phi} 
   \del_{\gammabar} 
   \period \label{currents}
\eqaend
Note that the last term in the second line. This does not affect the 
commutation relations but is necessary to assure the invariance of the
action. 

A convenient way to generate the primary fields is to use the 
following functionals \cite{Teschner1}
\eqabegin
   V^j &=& \Bigl[ (\gamma - x)(\gammabar - \xbar ) \, e^{\phi} 
   \, + \, e^{-\phi} \Bigr]^{2j}
  \comma \label{Vj}
\eqaend
where $x$ and $\xbar$ are some parameters.
By expanding $V^j$ in terms of $x^{j+m}$ and $\xbar^{j+\mbar}$, one gets 
the primary fields $V_{m,\bar{m}}^j $ with the definite eigenvalues
of $J_0^0 , \bar{J}_0^0 $ and the left and the right Casimirs.
It turns out that  $x$ and $\xbar$
are interpreted as the coordinates of the boundary CFT \cite{dBORT}.

Similarly to the Lorentzian case, the Hilbert space is decomposed into the 
representations of $SL(2,C)$. Schematically \cite{Gawedzki},
\eqabegin
  L^2 (H_3^+) & \sim & \int_0^\infty d \rho \, \rho^2 \Dpc^{-1/2+i\rho}
  \period
\eqaend  
We remark that only  
the principal continuous series $\Dpc$ appear
and there are no discrete series.\footnote{In section 2, we considered 
the representations of $SL(2,R)$. Here we are considering the corresponding
representations of $SL(2,C)$. In this case, the spectrum 
for $j=-1/2 + i \rho$ is given by $m= (ip+n)/2, \bar{m} = (ip-n)/2$ with 
$p \in \bfR \comma n \in \bfZ$.}

Furthermore, by (i) introducing auxiliary fields $\beta$ and 
$\betabar$, (ii) taking into account the non-trivial measure (\ref{measure})
and (iii) rescaling $\phi$, one obtains the following action,
\eqabegin
   S &=& \frac{1}{2\pi} \int d^2 \sigma \, \biggl(\, \del\phi \del \phi 
    + \beta \delbar \gamma + \betabar \del \gammabar - \beta \betabar \, 
   e^{-2\phi/\alpha_+} - \frac{2}{\alpha_+} \phi \sqrt{ \hat{g} } \hat{R} 
   \, \biggr) \comma  
\eqaend
where $\alpha_+ = \sqrt{2(k-2)} $ and $\hat{g}$ and $\hat{R}$ are the 
background metric and curvature of the world-sheet, respectively. 
In this expression, 
the interaction term $ \beta \betabar \, e^{-2\phi/\alpha_+} $ drops out
in the limit $\phi \to \infty $.
Thus we obtain a free theory in that limit, namely, near the boundary 
of $H_3^+$. The last term in $J^+_0$ in (\ref{currents})
also drops out and we get the free-field expression in that limit. 
\par\bigskip\noindent
{\it Puzzles}
\par\medskip

This WZW model has been intensively studied recently 
\cite{GKS}-\cite{Teschner2}
in relation to the AdS/CFT correspondence. We may need to be 
careful about to what extent and how the string theory without RR charges 
is relevant to the AdS/CFT correspondence (see, for example, \cite{SW}).
However, if the correspondence is naively taken, 
one  finds some puzzles. They are summarized 
in the following table of the correspondence,
\par\bigskip

\begin{center}
\begin{tabular}{ccc}
    string (WZW model) & \quad supergravity \quad & CFT \\ \hline\hline
    \quad 
   discrete series (non-normalizable) \quad & KK mode 
     & \quad chiral primary \quad \\
   \quad  \hspace{0.3em} 
   continuous series (normalizable) \quad \quad   &  ?? & ?? 
\end{tabular}
\end{center}
\par\bigskip
\noindent  
Namely, although the Hilbert space of the $H_3^+$ WZW model consists of the
principal continuous series, we do not find the corresponding objects 
on  the supergravity and the CFT sides. Moreover, 
following the argument in \cite{GKS}, the scaling dimension of 
a boundary CFT operator and the $sl_2$ spin of 
the corresponding operator of the $H_3^+$ WZW model are related by 
$ h = -j$.
If this is valid also for the principal continuous series, 
the dimension of the corresponding boundary CFT operator becomes complex.
Thus it is hard to interpret the correspondence for the continuous series
even if it exists.

These puzzles might not lead to an immediate contradiction because, 
as discussed in 
\cite{dBORT}, the $H_3^+$ WZW model is a non-compact CFT and hence
there might not be the state-operator correspondence as in the Liouville
theory. Nevertheless, in order to complete the correspondence, 
it seems necessary to further investigate these puzzles.
To this end, we may need to study the $H_3^+$ WZW model in detail.
Again, the fundamental consistency conditions play the role of 
the guideline there.

How are then the precise discussions of the $H_3^+$ WZW model possible?
One way is to use the free field approximation. By this approach, 
we can get much information \cite{GKS,GN} but this is valid only 
near the boundary 
$\phi \to \infty$. Another way is to use the generating functionals  
of the primary fields (\ref{Vj}) following 
\cite{Teschner1,dBORT,KS,Teschner2}. In this approach, the full analysis
beyond the free field approximation is possible but it tends to be 
semi-classical (see, however, \cite{Teschner1, Teschner2} regarding 
the full quantum analysis based on the bootstrap). 
Therefore it would be nice to have a description beyond the free field
or the semi-classical treatment. We will return to this point later.
\par \vskip 3ex\noindent
\mysection{Modular invariance}  
%
%

In the preceding sections, we put an emphasis on the importance of
the further investigations of the fundamental problems. Here we would 
like to discuss some attempts at clarifying the modular invariance 
of the $SL(2,R)$ WZW model. Although the modular invariance
 in the non-compact case
is not well understood, there are several arguments in the $SL(2,R)$
case. For example, the modular invariants are discussed in \cite{HHRS}
by using the $\hat{sl}(2,R)$ characters based on the discrete series 
$\Dhw$ and $\Dlw$ 
and by incorporating some new sectors corresponding to winding modes.
They are also discussed in \cite{YS} using the characters for $\Dpc$ 
along the line of \cite{Bars}.

In this section, we will focus on the possibility of constructing 
the modular invariants from
the characters for $\Dhw$ and $\Dlw$ without incorporating any additional
sectors as in \cite{HHRS}. For details, see \cite{KYS}. This issue 
is also discussed in \cite{PMP}.

Let us start with the definition of the characters.
For the current algebras based on  
compact Lie groups, the characters are naturally defined
using three variables. With this in mind, we define the characters for 
the discrete series by
\eqabegin
  {\rm ch}_j (z,\tau,u) & \equiv & e^{2\pi i k u} \sum e^{-2 \pi i J^0_0 z}
   e^{2 \pi i \tau (L_0 - \frac{c}{24})}
  \period \label{chj}
\eqaend
The summation is taken over the entire module of the current algebra 
representations. The plus sign in the first factor $ e^{+2 \pi i k u } $
is due to the change $ k \to - k $ compared with the compact case.
To calculate these characters, one needs to know about singular vectors.
For a generic highest (or lowest) weight representation, which is 
not necessarily $\Dhw$ or $\Dlw$, 
the current module has singular vectors when  
one of the following conditions is satisfied \cite{singularvec}:
\eqabegin
   (1) && 2 j + 1 = s + (k-2)(r-1) \comma \nn \\
   (2) && 2 j + 1 =  -s - r (k-2) \comma \label{singvec} \\
   (3) && k - 2 = 0  \comma \nn
\eqaend 
where $r,s$ are positive integers. 

The characters $ {\rm ch}_j$ in generic cases seem unknown. However,  
when there are no singular vectors, they are given by \cite{character,HHRS}
\eqabegin
   \chi^{\rm hw}_\mu (z,\tau,u) &=& e^{2 \pi i k u}
    e^{-2 \pi i \mu  z} q^{-\frac{ \mu^2}{k-2}} \ 
   i \vartheta_1^{-1}(z \vert \tau) 
    \comma \label{hwchi}
\eqaend
for $\Dhw$ and 
$ \chi^{\rm lw}_{\mu} (z,\tau,u) = \chi^{\rm hw}_\mu (-z,\tau,u)$ for 
$\Dlw$. 
Here $ q = e^{2 \pi i \tau}$, $ \mu \equiv j + 1/2 $ and 
\eqabegin 
   \vartheta_1 (z \vert \tau) & = & 2 q^{1/8} \sin (\pi z) 
  \prod_{n=1}^\infty (1-q^n)(1-q^n e^{2\pi i z})(1-q^n e^{-2\pi i z}) 
  \period \label{theta1}
\eqaend
Since $\chi^{\rm lw}_{\mu} (z,\tau,u) = - \chi^{\rm hw}_{-\mu} (z,\tau,u) $, 
$ \chi^{\rm hw}_{\mu}$ with $ \mu \geq  0 $ $(j \geq -1/2) $ are regarded 
as $-\chi^{\rm lw}_\mu $ with $ \mu \leq 0$. Thus we will 
use only $\chi^{\rm hw}_{\mu}$ and drop the superscript hw.
We remark that one cannot consider the specialized characters 
$\chi_\mu (0, \tau, 0)$ since they diverge in the limit $z \to 0$
because of the infinite degeneracy with respect to $L_0$.

In our normalization of $(z,\tau,u)$, 
the modular transformations are generated by  
\eqabegin
        S :  && (z, \tau, u ) \quad \to \quad     
                   \Bigl( \, \frac{z}{\tau}, -\frac{1}{\tau}, 
   u + \frac{z^2}{4\tau} \, \Bigr) \comma \nn \\
        T :  && (z, \tau, u ) \quad \to \quad (z, \tau + 1, u) 
    \period
\eqaend
Under $T$-transformation, the characters just get phases, 
\eqabegin
   \chi_\mu (z, \tau + 1, u) &=& 
   e^{-2 \pi i \lb \frac{\mu^2}{k-2} + \frac{1}{8}\rb }
    \chi_\mu (z, \tau, u)
 \period \label{T}
\eqaend
For $ k-2 < 0 $, the $S$-transformation of $\chi_\mu (z,\tau,0) $
is given in \cite{PMP}. In our case with three variables, it reads
as
\eqabegin
  \chi_\mu \Bigl( \,  \frac{z}{\tau}, -\frac{1}{\tau}, 
   u + \frac{z^2}{4\tau} \, \Bigr) & = &
  \sqrt{\frac{-2}{2-k} } \int_{-\infty}^\infty d\nu 
  \ e^{4 \pi i \frac{\mu \nu}{k-2}} \chi_\nu (z,\tau,u)
  \period \label{S1}
\eqaend
For $k-2 >0 $, the right-hand side of (\ref{S1}) does not converge 
on the upper half plane of $\tau$. 
Instead, after some calculation, we get a slightly different result,
\eqabegin
  \chi_\mu \Bigl( \, \frac{z}{\tau}, -\frac{1}{\tau}, 
   u + \frac{z^2}{4\tau} \, \Bigr) &=&
  \sqrt{\frac{-2}{k-2} } \int_{-\infty}^\infty d\nu 
  \ e^{- 4 \pi \frac{\mu \nu}{k-2}} \chi_{i\nu} (z,\tau,u)
  \period \label{S2}
\eqaend
Note that an imaginary $\mu = j + 1/2 $ corresponds to a spin 
of the principal continuous series but $\chi_{i\mu}$ are { not}
the characters for those representations.

As a simpler case, we will first discuss the possibility of 
constructing modular invariants using finite number of 
the discrete series characters.   
Given the explicit forms of $\chi_\mu$,
we can then show that {\it it is impossible to make modular invariants
from finite number of the discrete series characters without singular 
vectors, i.e., from $\chi_\mu $}.
Similarly, since the characters with singular vectors 
are obtained by subtracting states from $\chi_\mu$, the 
above statement is extended to some cases including singular vectors. 
In fact, we can show 
that, {\it  for $k > 2$, it is impossible to construct modular invariants
from finite number of the characters based on either $\Dhw$ or $\Dlw$}.
The arguments are simple applications of Cardy's for $c>1$ CFT 
\cite{Cardy} and we will omit them. To further extend the latter statement
to the cases including both $\Dhw$ and $\Dlw$, the explicit forms 
of the characters with singular vectors seem to be necessary.
In addition, we notice that a similar statement does not
hold for $k<2$. To see this, we note that the arguments 
do not use any special properties of the discrete unitary series 
and hence it is the same as for a generic 
highest (or lowest) weight $\hat{sl}(2,R)$ representations. However, 
when $k<2$, modular invariants using finite number of the characters 
are actually known for the so-called admissible representations 
\cite{admissible}.
 
Next, let us move on to the case in which infinitely many characters are 
allowed. For the time being, we will discuss the modular invariants
using $\chi_\mu$ only.
In such a case, using their modular properties we can show that 
{\it it is impossible to construct modular invariants only 
from $\chi_\mu$ with $\mu $ belonging to a finite interval 
$ \mu \in [ \mu_1 , \mu_2 ]$ even if infinitely many $\chi_\mu$ are used}.
This might seem obvious from the $S$-transformation of $\chi_\mu$
since the right-hand sides of (\ref{S1}) and (\ref{S2}) does not 
close within $ \chi_\mu $ with $ \mu \in [ \mu_1 , \mu_2 ]$. 
However, we need to be careful because we are considering
an infinite dimensional space of the characters $\chi_\mu$.
For instance, it is not clear which $\chi_\mu$ are independent 
and whether or not 
the expressions (\ref{S1}) and (\ref{S2}) are unique.   
In fact, it may be possible to get different expressions by deforming 
the integration contours in (\ref{S1}) and (\ref{S2}).
In any case, the detailed argument is given in \cite{KYS}. 

Since, for $k>2$,  $ \chi_\mu $ become divergent for $ \Im{ \tau } > 0 $ 
as $ \abs{ \mu } \to \infty $, the above statement means that 
{\it for $k>2$ it is impossible to construct modular invariants 
only from $\chi_\mu $}. 
Thus the possibility of constructing modular invariants from $\chi_\mu$ 
is limited to the case where $k<2$ and $\chi_\mu$ with $ \abs{ \mu } 
\to \infty$ are included. In this case, we can actually construct a 
modular invariant,
\eqabegin
   Z_{\rm diag} (z,\tau,u) &=& \int_{-\infty}^{\infty} d \mu \ 
   \abs{ \chi_\mu } ^2
      \ = \ \int_{-\infty}^{0} d\mu \ \bigl( \abs{ \chi^{\rm hw}_\mu } ^2 + 
             \abs{ \chi^{\rm lw}_\mu } ^2  \bigr) \nn \\
  &=& \half e^{- 4 \pi k \Im{ u } } 
  e^{(2-k)\pi \frac{(\Im{ z } )^2}{\Im{ \tau } } }
  \sqrt{ \frac{2-k}{ \Im{ \tau } } } 
  \abs{ \, \vartheta^{-2} (z\vert \tau) \, }
  \period 
\eqaend 
The diagonal partition function with $u = 0$, i.e., $Z_{\rm diag}(z,\tau,0)$, 
was discussed in \cite{PMP}. In our case, 
it is straightforward to check that $Z_{\rm diag}(z,\tau,u)$ is 
modular invariant owing to the presence of $u$.
Although it may be interpreted as a kind of a twisted partition function,
its  physical meaning is still unclear 
(recall that we cannot set $u = z= 0$).

As pointed out also in \cite{PMP}, 
$Z_{\rm diag}(z,\tau,0)$ was discussed 
in \cite{Gawedzki} in the context of a path-integral
approach to the $H_3^+$ WZW model. Since this model has the  
$\hat{sl}(2,C) \times \hat{sl}(2,C)^*$ symmetry, 
the diagonal partition function may be understood also 
as the partition function of this model. However, in \cite{Gawedzki}
different spectrum seems to be  summed up. It is interesting to
consider the precise relationship between the approach here and
the one in \cite{Gawedzki}.  

In order to discuss a generic case including the characters with 
singular vectors, we may again need  the explicit forms of such characters.  
Nevertheless, it turns out that the case without singular vectors covers 
physically interesting cases and gives important implication to 
the unitarity bound (\ref{ub}). This is because the condition 
of the singular vectors (\ref{singvec}) implies that there are 
no singular vectors within (\ref{ub}).
Furthermore, since the spins $j$ in that bound belong to a finite 
interval, our results indicate  that {\it 
one cannot construct modular invariants only from the discrete 
series characters based on the representations satisfying 
the unitarity bound (\ref{ub})}. This means that one cannot make 
a consistent string theory on $SL(2,R) = AdS_3$ only from the spectrum
within (\ref{ub}). This was already discussed in \cite{PMP}, but 
we believe that at least we have refined the argument a little. 

Since there exist ghosts for the discrete series outside the unitarity bound,
simply adding such spectrum may not give a consistent theory.  
Therefore, the possibilities for a consistent theory seem 
(a) to use  the discrete series satisfying (\ref{ub}) 
but include some new sectors with different characters from 
$\chi_\mu$ as in \cite{HHRS}, and/or
(b) to use the spectrum of other representations as in \cite{Bars,YS}.
To settle down this problem, further investigations are necessary.
%
%
\newpage
\mysection{Correlation functions}
%
%

In the previous section, we discussed the modular invariance 
and saw that it gives an important information about the spectrum
of the strings on $SL(2,R)$. Finally in this section, we will discuss
the calculation of the correlations functions of the $H_3^+$ WZW
model \cite{IOS}. This is important not only by itself but also
for studying the OPE and hence the spectrum of the model. 

The outstanding feature of the $H_3^+$ WZW model is that it allows us  
the Lagrangian approach \cite{Haba,Gawedzki}. It is alternative
and complementary to the current algebra approach which is often used.
Actually in the Lagrangian approach we may be able to get a description
beyond the free field and the semi-classical approximations.

To see this, let us first recall the action (\ref{H3+}) and 
the functional measure (\ref{measure}). Surprisingly, with (\ref{H3+})
and (\ref{measure}) it is possible to
carry out the path-integrals for some correlators \cite{Haba,Gawedzki}.
For this purpose, we need (i) the `partition function' obtained 
after integrating out $\gamma $ and $\gammabar$,
\eqabegin
   \exp \Bigl[ - S(\phi) \Bigr] &=& 
  \exp \biggl[ \,  \frac{1}{\pi} \int d^2  \sigma
  \Bigl( \, (k-2) \del \phi \del \phi 
  - \frac{1}{4} \phi \sqrt{\hat{g}} \hat{R} \, \Bigr) \, \biggr]
     \comma \label{Sphi} 
\eqaend
and (ii) the propagator,
\eqabegin
   \langle \, \gamma(z) \gammabar(\bar{w})\, \rangle
   &=& \frac{1}{k\pi} \int d^2y \, \frac{e^{-2\phi(y)}}{(z-y)(\bar{w}-\bar{y})}
  \period \label{prop}
\eqaend
The `partition function' 
implies that the resulting effective theory of $\phi$
is a free theory with a background charge. 
Since the the propagator contains the factor $e^{-2\phi(y)}$, 
it plays a similar role to the screening charges in the free field approach.
Using (\ref{Sphi}) and (\ref{prop}), one can calculate the correlators 
of the form
\eqabegin
   \Bigm\langle \prod_i \, e^{a_i \phi (z_i)} \gamma^{b_i} (z_i)
      \gammabar^{c_i} (\zbar_i) \Bigm\rangle
  \period
\eqaend

We are interested in the correlation functions of the primary
fields in the discrete series. 
In the free field approximation, the primaries  are given by
\eqabegin
   V^{j({\rm free})}_{m, \bar{m}} &=& 
   e^{2j\phi} \gamma^{j+m} \gammabar^{j+\mbar}
  \period
\eqaend
In the full theory, they are obtained 
by expanding the functionals (\ref{Vj}) and  take the form
\eqabegin
   V^j_{m,\mbar} &=& \sum_{j',m',\mbar'} C^{j'}_{m',\mbar'}
   V^{j'({\rm free})}_{m', \bar{m}'}
  \comma \label{fullVjmm}
\eqaend
where $C^{j'}_{m',\mbar'}$ are some coefficients.

We would like to calculate the correlators among the above
primary fields. A similar calculation was actually carried out 
in the case of the finite dimensional representations \cite{Gawedzki}.
In our infinite dimensional case, it seems that we need to carefully 
choose the `conjugate' fields paired with the above primaries.  
Once we have obtained the correlation functions,
we can extract important information
about the $H_3^+$ WZW model and in turn this gives useful insights
into the AdS/CFT correspondence. 
We would like to report progress in this direction elsewhere  \cite{IOS}.
\par \vskip 3ex\noindent
\mysection{Summary}
%
%

In this talk, we first noted  that the string theory on $AdS_3$
is important in various respects besides in relation to the AdS/CFT
correspondence. We then reviewed the $SL(2,R)$ and the $H_3^+$ WZW
models which describe the string propagations 
on Lorentzian and Euclidean $AdS_3$, respectively. 
We saw that in spite of the recent intensive
studies there still remain open questions at the fundamental level.
An emphasis was put on the importance of further investigating
the fundamental problems such as the modular 
invariance, the closure of the OPE, the issue of the spectrum and
the calculation of the correlation functions.
With this in mind, we discussed some attempts at clarifying such problems. 
First,
we discussed the modular invariance of the $SL(2,R)$ WZW model and
showed that it gives important information of the spectrum.
Next, we discussed the correlation functions of the $H_3^+$
WZW model using the full Lagrangian approach. 

Although our attempts were quite incomplete, some of the problems seem
still tractable. Thus further investigations will lead us  to
a deeper understanding of the string theory on $AdS_3$.
We expect that such investigations also shed some light on 
the AdS/CFT correspondence.  
\vskip 7ex 
\noindent
{\bf Note added}
\par \medskip
Some of the questions raised in this talk have been discussed 
also in recent papers \cite{MOLS}.
%
\newpage
\begin{center}
 {\sc Acknowledgements}
\end{center}
\par \bigskip

I would like to thank N. Ishibashi, A. Kato and K. Okuyama for 
the collaborations on the subjects discussed here. I would also 
like to thank I. Bars, J. de Boer, A. Giveon, H. Ishikawa, K. Ito,  
M. Kato, P.M. Petropoulos and S.-K. Yang for useful discussions 
and correspondences. Finally, I am grateful to the organizers of the 
workshop `Developments in Superstring and M-theory' held at YITP,
Kyoto, 27-29 October, 1999, for giving  a chance to deepen my 
understanding on the string theory on $AdS_3$.
%
%
%
%
\def\thebibliography#1{\list
 {[\arabic{enumi}]}{\settowidth\labelwidth{[#1]}\leftmargin\labelwidth
  \advance\leftmargin\labelsep
  \usecounter{enumi}}
  \def\newblock{\hskip .11em plus .33em minus .07em}
  \sloppy\clubpenalty4000\widowpenalty4000
  \sfcode`\.=1000\relax}
 \let\endthebibliography=\endlist
\vskip 10ex
\begin{center}
 {\sc References}
\end{center}
\par \smallskip


\begin{thebibliography}{999}
\parskip=-1pt

%
\bibitem{BFOW} J. Balog, L. O'Raifeartaigh, P. Forg{\'a}cs and A. Wipf,
 \npb325(1989)225.
%
\bibitem{ng1} 
     P.M. Petropoulos, \plb236(1990)151; \\  
     N. Mohameddi, \ijmp5(1990)3201; \\
     S. Hwang, \npb354(1991)100; \\ 
     J. Evans, M. Gaberdiel and M. Perry, \npb535(1998)152, \xxx9806024.
%
\bibitem{HHRS}
   M. Henningson, S. Hwang, P. Roberts and B. Sundborg, \plb267(1991)350.
%
\bibitem{BN} I. Bars and D. Nemeschansky, \npb348(1991)89.
%
\bibitem{Bars} I. Bars,  \prd53(1996)3308, \xxx9503205.
%
\bibitem{YS} Y. Satoh, \npb513(1998)213, \xxx9705208.
%
\bibitem{Gawedzki} 
 K. Gaw\c{e}dzki, \npb328(1989)733; ~
 {\it Non-compact WZW conformal field theories}, \xxx9110076,  
   ~Proceedings of the Carg\`ese Summer Institute.
%
\bibitem{Teschner1} J. Teschner, \npb546(1999)390, \xxx9712256;
                   ~ \npb546(1999)369, \xxx9712258.
%
\bibitem{crvdstrng} J.G. Russo and A.A. Tseytlin, \npb449(1995)91; \\
      E. Kiritsis, C. Kounnas and D. L\"{u}st, \plb331(1994)321.
%
\bibitem{sl2/u1} E. Witten, \prd44(1991)314; \\ 
     G. Mandal, A. Senguputa and S. Wadia, \mpla6(1991)1685.
%
\bibitem{3dbh} G.T. Horowitz and D.L. Welch, 
         \prl71(1993)328, \xxx9302126 ; \\
          N. Kaloper, \prd48(1993)2598, \xxx9303007;\\
            A. Ali and A. Kumar, \mpla8(1993)2045, \xxx9303032.
%
\bibitem{NS}  M. Natsuume and Y. Satoh, \ijmp13(1998)1229, \xxx9611041.
%
\bibitem{BTZ}  M. Ba{\~ n}ados, C. Teitelboim and J. Zanelli, 
                  \prl69(1992)1849 (1992), \xxx9204099; \\ 
               M. Ba{\~ n}ados, M. Henneaux, C. Teitelboim and J. Zanelli,
                  \prd48(1993)1506, \grqc9302012.
%
\bibitem{HSS} S. Hyun, \xxx9704005, {\it U duality between three-dimensional
                and higher dimensional black holes}; \\
              K. Sfetsos and K. Skenderis, \npb517(1998)179, \xxx9711138. 
%
\bibitem{CT} M. Cveti{\v c} and A.A. Tseytlin, \plb366(1996)95.
%
\bibitem{YS2} Y. Satoh, \prd59(1999)084010, \xxx9810135.
%
\bibitem{AdSCFT} 
    J. Maldacena,  
      Adv. Theor. Math. Phys. {\bf 2} (1998) 231, \xxx9711200; \\
      S.S. Gubser, I.R. Klebanov and A.M. Polyakov, \plb428(1998)105,
     \xxx9802109; \\ 
      E. Witten, Adv. Theor. Math. Phys. {\bf 2} (1998) 2253, \xxx9802150.
%
\bibitem{GS} I. Pesando, JHEP  {\bf 9902} (1999) 7, \xxx9809145; \\
   J. Rahmfeld and A. Rajaraman,  \prd60(1999)064014, \xxx9809164;\\
   J. Park and S.-J. Rey, JHEP  {\bf 9901} (1999) 1, \xxx9812062.
%
\bibitem{BVW} N. Berkovits, C. Vafa and E. Witten, \jhep9903(1999)018,
              \xxx9902098.
%
\bibitem{GKS} 
      A. Giveon, D. Kutasov and N. Seiberg, 
               Adv. Theor. Math. Phys. {\bf 2} (1998) 733, \xxx9806194. 
\bibitem{dBORT}   J. de Boer, H. Ooguri, H. Robins and J. Tannenhauser, 
               JHEP {\bf 9812} (1998) 26, \xxx9812046.
\bibitem{SL2WZW}
      K. Ito, \plb449(1999)48, \xxx9811002; \\
      K. Hosomichi and Y. Sugawara, \jhep9901(1999)013, \xxx9812100; \\
      S. Yamaguchi, Y. Ishimoto and K. Sugiyama, \jhep9902(1999)026, 
   \xxx9902079.
\bibitem{KS}      
  D. Kutasov and N. Seiberg, JHEP {\bf 9904} (1999) 8, \xxx9903219.
%
\bibitem{Teschner2} J. Teschner,   {\it Operator product expansion 
           and factorization in the $H_+^3$ WZNW model}, \xxx9906215. 
%
\bibitem{BDM} I. Bars, C. Deliduman and D. Minic, 
        {\it String theory on $AdS_3$ revisited}, \xxx9907087.
%
\bibitem{PMP} P.M. Petropoulos, 
    {\it String theory on ADS$_3$: some open questions}, 
   \xxx9908189.
%
\bibitem{KYS} A. Kato and Y. Satoh, to appear 
({\it Modular invariance of string theory on $AdS_3$}, \xxx0001063).
%
\bibitem{IOS} N. Ishibashi, K. Okuyama and Y. Satoh, 
         work in progress 
   ({\it Path integral approach to string theory on $AdS_3$},
    \xxx0005152).
%
\bibitem{VK} N. Ja. Vilenkin and A.U. Klimyk, {\it Representations
             of Lie Groups and Special Functions}, 
         (Kluwer Academic Publishers, Dordrecht, 1991).
%
\bibitem{FT} E.S. Fradkin and A.A. Tseytlin, \npb271(1986)333.
%
\bibitem{GO} P. Goddard and D. Olive, \ijmp1(1986)303.
%
\bibitem{GW} D. Gepner and E. Witten, \npb278(1986)493.
%
\bibitem{Wakimoto} M. Wakimoto, \cmp104(1986)605.
%
\bibitem{SW} N. Seiberg and E. Witten, \jhep9904(1999)017, \xxx9903224.
%
\bibitem{GN} G. Giribet and C.A. N{\'u}{\~n}ez, \jhep9911(1999)031, 
   \xxx9909149.
%
%
%
\bibitem{singularvec} 
   V.G. Ka\v{c} and D.A. Kazhdan, Adv. Math. {\bf 34} (1979) 97; \\ 
   F.G. Malikov, B.L. Feigin and D.B. Fuks, 
      Funct. Anal. and Appl. {\bf 20} (1986) 103. 
%
\bibitem{character} 
  P.A. Griffin and O.F. Hernandez, \npb356(1991)287; \\
  K. Sfetsos, \plb271(1991)301; \\
  I. Bakas and E. Kiritsis, \ijmp7(1992)55, \xxx9109029; \\
  P.M. Petropoulos, Th\`ese de doctorat, Ecole Polytechnique, 1991.
%
%
\bibitem{Cardy} J. Cardy, \npb270(1986)186.
%
\bibitem{admissible} V. Ka\v{c} and M. Wakimoto, 
            Proc. Nat. Acad. Sci. USA {\bf 85} (1988) 4956; \\
     I.G. Koh and P. Sorba, \plb215(1988)723.
%
%
\bibitem{Haba} Z. Haba, \ijmp4(1989)267.
%
\bibitem{MOLS} 
     J. Maldacena and H. Ooguri, {\it Strings in $AdS_3$ and
          $SL(2,R)$ WZW model: I}, \xxx0001053; \\
     A.L. Larsen and N. S\'{a}nchez, {\it Quantum coherent string states
          in $AdS_3$ and $SL(2,R)$ WZWN model}, \xxx0001180; \\
     I. Pesando, {\it Some remarks on the free fields realization of 
       the bosonic string on $AdS_3$}, \xxx0003036; \\
     Y. Hikida, K. Hosomichi, Y. Sugawara, 
       {\it String theory on $AdS_3$ as discrete light-cone Liouville
        theory}, \xxx0005065.
%
\end{thebibliography}
\end{document}